\newcommand{\CLASSINPUTtoptextmargin}{0.78in}
\newcommand{\CLASSINPUTbottomtextmargin}{1.03in}

\documentclass[conference]{IEEEtran}
\IEEEoverridecommandlockouts
\usepackage{cite}
\usepackage{amsmath,amssymb,amsfonts}
\usepackage{algorithmic}
\usepackage{graphicx}
\usepackage{textcomp}
\usepackage{xcolor}
\usepackage{glossaries}
\usepackage{siunitx}
\usepackage[hidelinks]{hyperref}
\usepackage{textpos}

\newacronym{XR}{XR}{Extended Reality}
\newacronym{6DoF}{6DoF}{6 Degrees of Freedom}
\newacronym{AP}{AP}{Access Point}
\newacronym{SLS}{SLS}{Sector-Level Sweep}
\newacronym{DTI}{DTI}{Data Transmission Interval}
\newacronym{AWV}{AWV}{Antenna Weight Vector}
\newacronym{TXSS}{TXSS}{Transmit Sector Sweep}
\newacronym{A-BFT}{A-BFT}{Association - Beamforming Training}
\newacronym{VR}{VR}{Virtual Reality}
\newacronym{BHI}{BHI}{Beacon Header Interval}
\newacronym{BI}{BI}{Beacon Interval}
\newacronym{MCS}{MCS}{Modulation and Coding Scheme}
\newacronym{QoE}{QoE}{Quality of Experience}
\newacronym{CDF}{CDF}{Cumulative Distribution Function}
\newacronym{mmWave}{mmWave}{Millimeter-Wave}
\newacronym{BRP}{BRP}{Beam Refinement Phase}
\newacronym{BTI}{BTI}{Beacon Transmission Interval}
\newacronym{LoS}{LoS}{Line of Sight}
\newacronym{HR2LLC}{HR2LLC}{High Rate, High Reliability and Low Latency Communications}
\newacronym{MIMO}{MIMO}{Multiple-Input Multiple-Output}
\newacronym{EVK}{EVK}{Evaluation Kit}
\newacronym{RFIC}{RFIC}{Radio-Frequency Integrated Circuit}
\newacronym{NPU}{NPU}{Network Processor Unit}
\newacronym{CBAP}{CBAP}{Contention-Based Access Period}
\newacronym{SP}{SP}{Service Period}
\newacronym{PTP}{PTP}{Precise Time Protocol}
\newacronym{SSW}{SSW}{Sector Sweep}
\newacronym{AoA}{AoA}{Angle of Arrival}
\newacronym{SNR}{SNR}{Signal-to-Noise Ratio}
\newacronym{MAC}{MAC}{Medium Access Control}
\newacronym{FoV}{FoV}{Field of View}
\newacronym{RIS}{RIS}{Reconfigurable Intelligent Surface}
\newacronym{ATI}{ATI}{Announcement Transmission Interval}

\def\BibTeX{{\rm B\kern-.05em{\sc i\kern-.025em b}\kern-.08em
    T\kern-.1667em\lower.7ex\hbox{E}\kern-.125emX}}
\begin{document}

\title{Can Millimeter-Wave Support Interactive Extended Reality under Rapid Rotational Motion?
}

\author{

\IEEEauthorblockN{Jakob Struye\IEEEauthorrefmark{1}, Hany Assasa\IEEEauthorrefmark{2}, Barend van Liempd\IEEEauthorrefmark{2}, Arnout Diels\IEEEauthorrefmark{3}, Jeroen Famaey\IEEEauthorrefmark{1}\\
\IEEEauthorblockA{\IEEEauthorrefmark{1}University of Antwerp - imec, Antwerp, Belgium. Email: \{firstname\}.\{lastname\}@uantwerpen.be }
\IEEEauthorblockA{\IEEEauthorrefmark{2}Pharrowtech, Leuven, Belgium. Email: \{hany,barend.vanliempd\}@pharrowtech.com}
\IEEEauthorblockA{\IEEEauthorrefmark{3}Dekimo, Leuven , Belgium. Email: \{firstname\}.\{lastname\}@dekimo.com}
}}
\maketitle

\begin{abstract}
Using Millimeter-Wave (mmWave) wireless communications is often named as the prime enabler for mobile interactive Extended Reality (XR), as it offers multi-gigabit data rates at millisecond-range latency. To achieve this, mmWave nodes must focus their energy towards each other, which is especially challenging in XR scenarios, where the transceiver on the user's XR device may rotate rapidly. To evaluate the feasibility of mmWave XR, we present the first throughput and latency evaluation of state-of-the-art mmWave hardware under rapid rotational motion, for different PHY and MAC-layer parameter configurations. We show that this parameter configuration has a significant impact on performance, and that specialized beamforming approaches for rapid rotational motion may be necessary to enable uninterrupted, high-quality mobile interactive XR experiences.
\end{abstract}

\begin{textblock}{180}(0,7.8)
    \begin{tiny}
    \copyright\ 2024 IEEE. Personal use of this material is permitted. Permission from IEEE must be obtained for all other uses,\\
    \vspace{-6mm}\\
    in any current or future media, including reprinting/republishing this material for advertising or promotional purposes,\\
    \vspace{-6mm}\\
    creating new collective works, for resale or redistribution to servers or lists, or reuse of any copyrighted component\\
    \vspace{-6mm}\\
    of this work in other works.
  \end{tiny}
\end{textblock}
\section{Introduction}
\gls{XR} has found a wide array of applications over the past decade, including in healthcare, education, entertainment and manufacturing~\cite{applications}. The technical requirements of deployments depend heavily on the exact application. In the most challenging case, extremely-high-quality content is captured or generated in real-time, and transmitted wirelessly to \gls{XR}-enabled devices, such as head-mounted devices or smartphones. In interactive applications, which require real-time content acquisition, data rates are usually extremely high. Heavy-duty compression cannot be deployed to reduce data rate requirements, as these algorithms are time-consuming, which increases latency. Furthermore, compression is less effective when there is no option of looking ahead to upcoming content. In several types of applications, incorporating a wireless link is essential. This includes applications where content acquisition inherently occurs off-site (e.g., remote conferencing) or when off-loading rendering to an edge cloud enables high-quality virtual experiences even with small, light-weight and silent devices (e.g., gaming). When a satisfactory \gls{QoE} requires multi-gigabit data rates with latencies in the order of milliseconds along with near-perfect reliability~\cite{towardVR1}, sub-\SI{6}{\giga\hertz} wireless communications are no longer sufficient. Only higher frequencies offer the bandwidth necessary to fulfill these \gls{HR2LLC} requirements~\cite{HR2LLC}. Of these, \gls{mmWave}, spanning 24 to \SI{300}{\giga\hertz}, is most likely to be deployed at a large scale in the near future, as \gls{mmWave} devices aimed at consumers have been available for some years. Performance figures attainable with \gls{mmWave} are highly impressive, with the most recent \gls{mmWave} amendment for Wi-Fi, IEEE 802.11ay, offering a maximum link rate of over \SI{8.5}{Gbps} using a single \SI{2.16}{\giga\hertz} channel. Through channel aggregation and \gls{MIMO}, an \gls{AP} could theoretically reach over \SI{275}{Gbps}, divisible among multiple users. Realizing these performance levels does come with a number of challenges not prominent in sub-\SI{6}{\giga\hertz} bands. As path and penetration losses increase along with frequency, establishing and maintaining links with sufficiently high signal strength for communications at high \glspl{MCS} becomes highly challenging. The main mitigation technique for this is \textit{beamforming}, a process in which transceivers focus their energy in specific, carefully selected directions. In its most common implementation, the singular antenna is replaced with an \textit{antenna array} consisting of many carefully positioned antenna elements. By applying a different, intelligently selected, phase shift to each element's signal, the different signals will be phase-aligned, and therefore interfere constructively, in certain directions, but interfere destructively in others. Essentially, this enables focusing a large fraction of the available energy budget in certain desirable directions, effectively increasing the resulting signal strength significantly. One configuration of phase shifts is colloquially called a beam, and beamforming is commonly implemented through \textit{codebooks} containing a pre-defined list of beams. The optimal beam between two devices can be determined by exhaustively sweeping all beams in the codebook.

Despite the first version of the Wi-Fi \gls{mmWave} amendment, IEEE 802.11ad, being over a decade old, hardware availability and, as a result, performance evaluations, have remained limited. Several works evaluate mmWave-capable Dell hardware, with evaluations on throughput given different distance, \gls{AoA} and interference levels~\cite{delleval,dellonbot}, on reverse engineering the protocol configuration and beamforming effectiveness~\cite{boonbane}, and on power consumption~\cite{powerconsumption}. Also frequently evaluated is the TP-Link Talon AD7200, with evaluations on \gls{SNR} with different beams~\cite{talonbeams}, throughput under blockage and translational motion~\cite{fastinfuroating} and performance during mobility-induced blockage~\cite{talononbot}. Evaluations with the mmWave-equipped Asus ROG phone include those on throughput under translational motion~\cite{phoneeval} and on performance in multi-\gls{AP} scenarios~\cite{phonemulti}. Finally, several evaluations of experimental \SI{60}{\giga\hertz} testbeds investigate the impact of mobility on performance~\cite{X60eval,MMFLEX,MIMORPH}. Overall, no existing work evaluates throughput and latency \textit{during} rotational motion of an IEEE 802.11ad/ay system with tunable PHY and \gls{MAC}-layer parameters. In this work, we evaluate the performance of a state-of-the-art \gls{mmWave} \textit{\gls{EVK}}, with a focus on its performance for \gls{XR}. Specifically, we evaluate throughput and latency under mobility, and evaluate the impact of changing system parameters on performance.

\begin{figure}[!t]
  \centering
  \includegraphics[width=\linewidth]{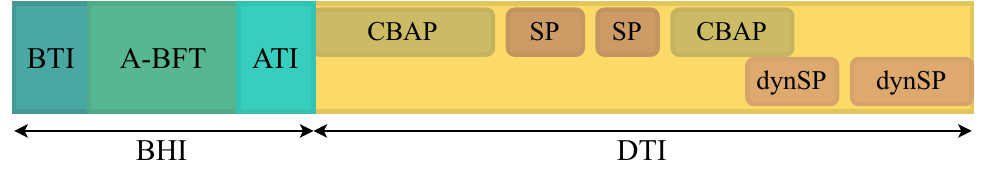}
  \caption{The Beacon Interval}
  \label{fig:biconfig}
\end{figure}
\section{IEEE 802.11ad/ay}
This work considers hardware implementing the IEEE 802.11ad protocol along with its backwards-compatible successor IEEE 802.11ay, both incorporating \gls{mmWave} functionality into Wi-Fi. In this section, we provide protocol details along with an overview of our hardware implementing it.
\subsection{Protocol details}
As shown in Fig.~\ref{fig:biconfig}, IEEE 802.11ad/ay divides its transmission schedule into \glspl{BI} of configurable length~\cite{80211adoverviewnitsche}. Each \gls{BI} starts with a \gls{BHI} reserved for overheads. The \gls{BHI} starts with a \gls{BTI} during which the \gls{AP} transmits a beacon advertising its existence and capabilities using its different beams. Then, during the \gls{A-BFT} phase, consisting of different slots, devices may attempt to associate to the \gls{AP}. This is followed by the optional \gls{ATI} for additional signalling. After the \gls{BHI}, there is a \gls{DTI}, intended for actual data transmission. The \gls{AP} orchestrates the \gls{DTI}, subdividing it into periods. Each period is either a \gls{CBAP}, during which any node may transmit after a backoff period, or a \gls{SP}, which is reserved for communications between two specific nodes.
\subsection{Hardware}
These experiments use two \glspl{EVK} supplied by Pharrowtech, configured as \gls{AP} and client respectively. The \gls{EVK} incorporates Pharrowtech's SPIRIT PTM1060 module, containing the \gls{RFIC} and antenna array, with the Renesas RWM6050 baseband processor. The \gls{EVK} also contains an Intel NUC serving as \gls{NPU}, running Linux, through which users can configure the other components. The \gls{EVK} complies with the IEEE 802.11ad/ay standard, and covers its full unlicensed spectrum range, from 57 to \SI{71}{\giga\hertz}. It supports channel bonding of two channels, 64-QAM modulation, an \gls{MCS} up to 9, and high-resolution phase-shifting.
\section{Experiments}
In these experiments, we deploy one \gls{EVK} as a static \gls{AP} and another as a mobile user. This mimics a single-user interactive \gls{XR} secnario, where the user may roam freely within a room, and the \gls{AP} can cover it throughout the room. During the evaluations, both \glspl{EVK} were positioned on a table, with rotations being performed manually. The two devices were placed \SI{3}{\meter} apart horizontally, which is within the range expected in \gls{XR} deployments. Unless noted differently, we configure the devices to use \gls{MCS} 9, enable full MAC-layer aggregation, set the transmit power to levels appropriate for indoor usage and a \gls{BI} of \SI{10}{\milli\second}. Latency is measured using our open-source \texttt{bwdelaytester}\footnote{\url{https://github.com/arnoutdekimo/bwdelaytester}} tool, configured to send \SI{1000}{\byte} packets using UDP at \SI{1.6}{Gbps}, which does not sature the link using the default configuration. At the start of each experiment, the \glspl{EVK}' clocks are synchronized with sub-microsecond accuracy using \gls{PTP}, and each packet contains a generation timestamp, meaning per-packet latency can be measured extremely accurately. We visualize the packet latency with a \gls{CDF}, and include the packet loss in each plot. Note that, given the real-time nature of the application, we considered packets with a latency over \SI{10}{\milli\second} to also be lost. Each latency experiment is run for 4 minutes. In some scenarios, we also report the maximal throughput, which was measured in a separate experiment, using saturated UDP \texttt{iPerf}\footnote{\url{https://iperf.fr}} traffic.

In this evaluation, we vary parameters one by one. We investigate the impact of different \gls{MCS} settings, \gls{BI} lengths, \gls{BHI} configurations, channel access schemes, and beam tracking\footnote{We use ``beam tracking'' to refer to beamforming when it occurs for an already-associated mobile device} approaches. To isolate the impact of each parameter, we perform this first set of experiments in a static scenario, without rotating the user \gls{EVK}. In a second step, we repeat a subset of well-performing configurations under mobility. We consider three different mobility levels: in the first two, the user performs either moderate or rapid rotational motion while the \gls{AP} remains within its \gls{FoV}, while for the third level, the user \gls{EVK} is rotated far enough for the \gls{AP} to move outside of its \gls{FoV}.

\section{Results}
In this section, we present the results of each experiment, considering both static and mobile deployments.
\begin{figure}[!t]
  \centering
  \includegraphics[width=\linewidth]{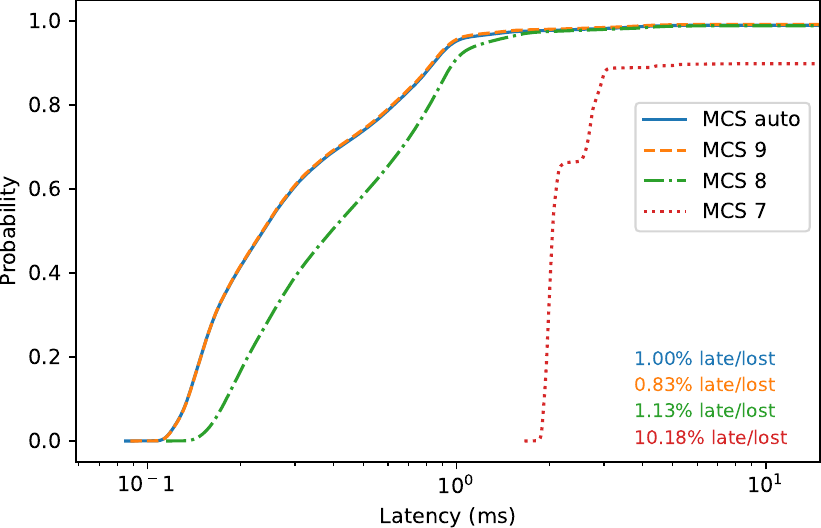}
  \caption{Latency for different \gls{MCS} settings}
  \label{fig:mcs}
\end{figure}
\begin{figure}[!t]
  \centering
  \includegraphics[width=\linewidth]{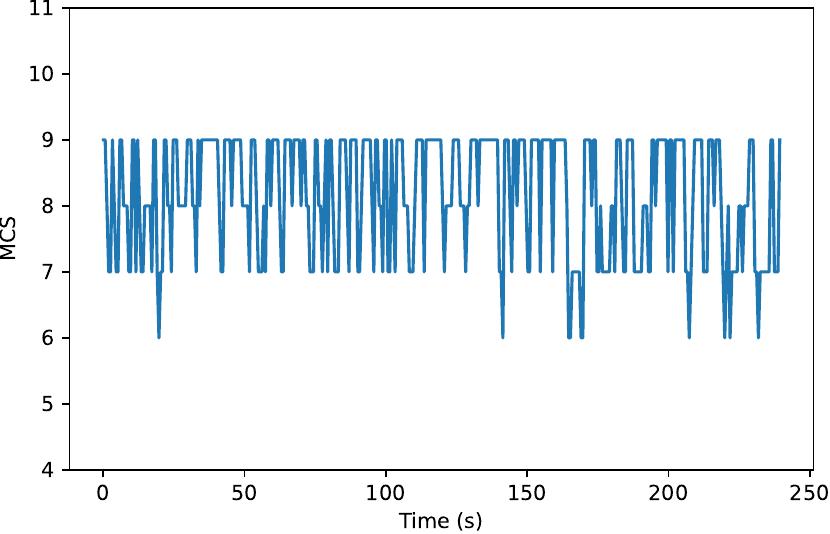}
  \caption{Actual \gls{MCS} selected by rate adaptation}
  \label{fig:mcs_adapt}
\end{figure}
\subsection{Static Experiments}
First, we consider \textbf{\gls{MCS} tuning}. By default, the \gls{MCS} is at 9, the highest supported value. We also evaluate lower settings, along with automatic rate adaptation. For static indoor scenarios, \gls{MCS} 9 is easily decodable, meaning we expect a reduction in performance for lower \gls{MCS} and negligible impact from enabling automatic rate adaptation. We first evaluate the throughput, being \SI{1.85}{Gbps} for \gls{MCS} 9 (and rate adaptation), \SI{1.73}{Gbps} for \gls{MCS} 8 and \SI{1.45}{Gbps} for \gls{MCS} 7. As we evaluate latency at \SI{1.6}{Gbps}, we do not lower \gls{MCS} any further. Fig.~\ref{fig:mcs} shows the results of the latency experiments. At \gls{MCS} 7, the data rate inflates the latency, and \gls{MCS} 8 still leads to higher latency than \gls{MCS} 9, due to increased transmission delay at lower throughput. Note that most experiments exhibit a loss between \SI{0.5}{\percent} and \SI{1}{\percent}, which we investigate throughout this section. For rate adaptation, Fig.~\ref{fig:mcs_adapt} visualizes the actual \gls{MCS} selected through time, showing that it mostly fluctuates between 7 and 9, with a few brief reductions to 6. We run the remaining experiments at \gls{MCS} 9 to eliminate the adaptation algorithm as a variable.
\begin{figure}[!t]
  \centering
  \includegraphics[width=\linewidth]{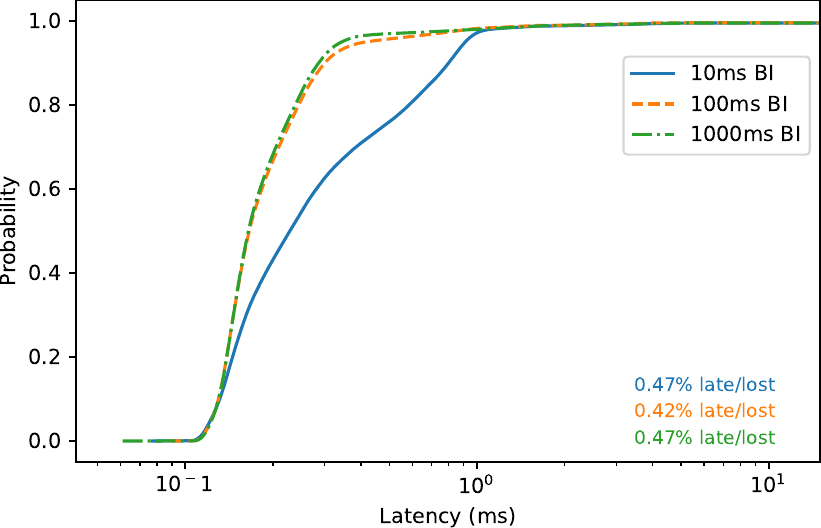}
  \caption{Latency for different \gls{BI} lengths}
  \label{fig:bi}
\end{figure}

Next, we evaluate the impact of the \textbf{\gls{BI} length}, which crucially determines the time between \glspl{BHI}. During each \gls{BHI}, no data transmission can occur, meaning that increasing the \gls{BI} length should reduce latency. On the other hand, fewer \glspl{BHI} may lead to performance degradation, as this period may be used for beam tracking. However, the default configuration can beam track in the \gls{DTI}, meaning we expect this impact to be limited. We experiment with \SI{10}{\milli\second}, \SI{100}{\milli\second} and \SI{1000}{\milli\second} \glspl{BI}, with results in Fig.~\ref{fig:bi}. Clearly, latency is slightly lower with fewer \glspl{BHI}. This makes sense, as packets are briefly blocked during the \gls{BHI}, so more packets will experience this with more frequent \glspl{BHI}. The three \glspl{CDF} converge around the \SI{1}{\milli\second} mark, which is roughly the duration of the \gls{BHI}. The throughput, instead, experiences a considerable increase when going from \SI{10}{\milli\second} to \SI{100}{\milli\second}, jumping from \SI{1.85}{Gbps} to \SI{1.98}{Gbps}. The further increase to \SI{1000}{\milli\second} only improves throughput by another \SI{0.01}{Gbps}. This makes sense, as the absolute reduction in scheduled \gls{BHI} time is only one tenth of what it was when going from \SI{10}{\milli\second} to \SI{100}{\milli\second}.
\begin{figure}[!t]
  \centering
  \includegraphics[width=\linewidth]{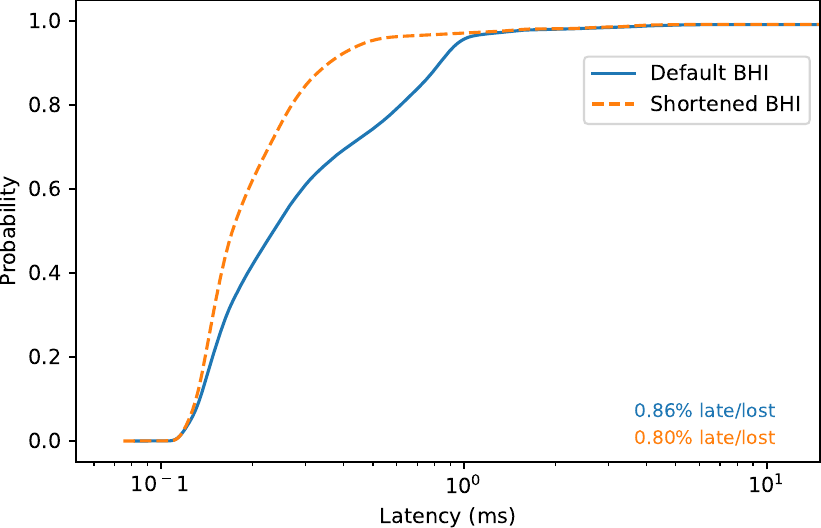}
  \caption{Latency for different \gls{BHI} configurations (\SI{10}{\milli\second } \glspl{BI})}
  \label{fig:bhilen}
\end{figure}

By default, the \gls{BHI} is configured to contain 8 beacons and 1 \gls{SSW} slot of 16 frames in the \gls{A-BFT} phase. The specification requires at least 1 beacon per \gls{BHI}, 1 frame per \gls{SSW} slot, and an \gls{A-BFT} only once every 15 \glspl{BHI}. Using this \textbf{shortened \gls{BHI} configuration} is expected to improve performance in a static scenario~\cite{towards}. At \SI{10}{\milli\second}, this results in the latencies shown in Fig.~\ref{fig:bhilen}, measured with all optimizations enabled. Similar to the previous experiment, the \gls{CDF} rises more quickly with the shorter \glspl{BHI}, as blocked packets are released more rapidly with shorter \gls{BHI}. At higher \gls{BI} values (100 to \SI{1000}{\milli\second}) these optimizations had little to no impact, as the fraction of time reserved for \glspl{BHI} is already significantly lower. Each optimization separately improves throughput by 0.04 to \SI{0.07}{Gbps}. All three enabled together results in only a \SI{0.11}{Gbps} increase, as reducing the length of the \gls{A-BFT} is less impactful when also reducing its frequency.
\begin{figure}[!t]
  \centering
  \includegraphics[width=\linewidth]{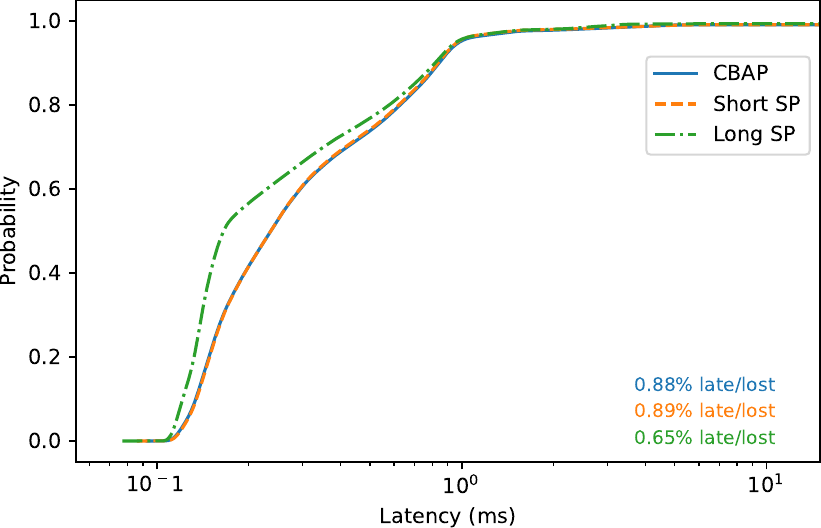}
  \caption{Latency for different channel access approaches (\SI{10}{\milli\second} BI)}
  \label{fig:sched}
\end{figure}
\begin{figure}[!t]
  \centering
  \includegraphics[width=\linewidth]{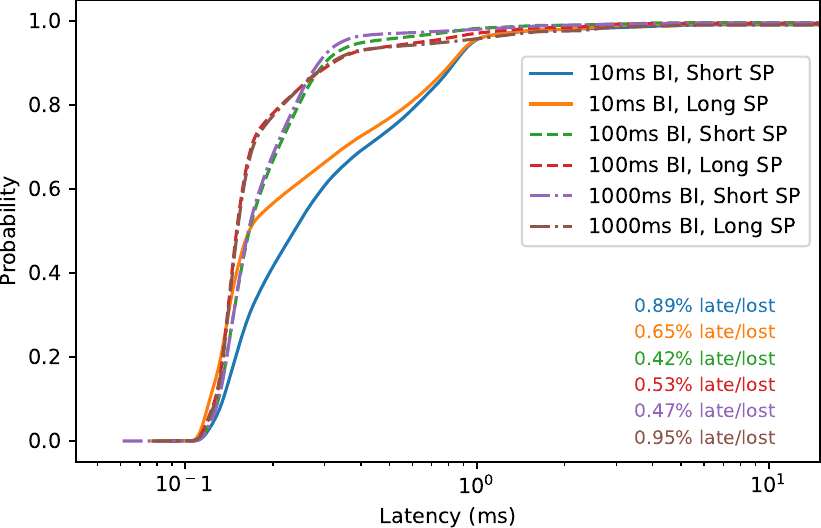}
  \caption{Latency for short vs long \gls{SP} at different \gls{BI} lengths.}
  \label{fig:sched_bi}
\end{figure}

\textbf{Channel access} can be configured with either \glspl{CBAP} or \glspl{SP}. As the \glspl{EVK} implement \glspl{CBAP} on top of \glspl{SP}, and as most transmissions are from the \gls{AP} (which does not need to request \glspl{SP}) we assume the configuration will make little difference. However, with \glspl{SP} there is also a parameter which controls the maximal \gls{SP} length. As there are buffer periods between subsequent \glspl{SP}, increasing this may improve performance. By default, the \gls{SP} length is \SI{2}{\milli\second}, with a maximum allowed value of \SI{65}{\milli\second}. Fig.~\ref{fig:sched} shows that, as assumed, the difference between \gls{CBAP} and default \gls{SP} is negligible, indicating the \gls{CBAP} is implemented using a \SI{2}{\milli\second} \gls{SP}. The impact of the longer \glspl{SP} is noticeable but limited. However, this was run at a \gls{BI} of \SI{10}{\milli\second}, meaning the full \SI{65}{\milli\second} could not be reached. We therefore repeat the \gls{SP} measurements at \SI{100}{\milli\second} and \SI{1000}{\milli\second} \glspl{BI}. We compare this to the \SI{10}{\milli\second} long \gls{SP} configuration in Fig.~\ref{fig:sched_bi}. While long \glspl{SP} perform consistently better latency-wise at \SI{10}{\milli\second} \gls{BI}, this is no longer the case for higher \gls{BI} values: the short \gls{SP} \gls{CDF} eventually overtakes the long \gls{SP}’s. There may be an optimal configuration somewhere between \SI{2}{\milli\second} and \SI{65}{\milli\second}. We leave optimizing this for future work, as the potential gains appear limited. In addition, we note that these parameters affect throughput significantly: lengthening either the \glspl{SP} to \SI{65}{\milli\second} or \gls{BI} to \SI{100}{\milli\second} gains \SI{0.12}{Gbps} and \SI{0.13}{Gbps} respectively, with the combination gaining \SI{0.27}{Gbps}, as this allows for the full \SI{65}{\milli\second} \gls{SP} to be scheduled. Increasing the \gls{BI} further to \SI{1000}{\milli\second} did not markedly improve latency, but did lead to significant fluctuations in throughput.
\begin{figure}[!t]
  \centering
  \includegraphics[width=\linewidth]{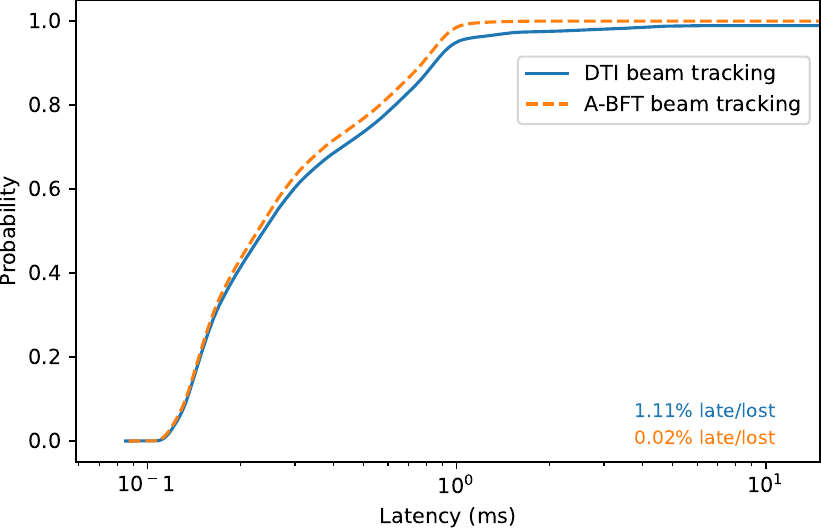}
  \caption{Latency for beam tracking in different periods (\SI{10}{\milli\second } \glspl{BI})}
  \label{fig:dtibf}
\end{figure}
\begin{figure}[!t]
  \centering
  \includegraphics[width=\linewidth]{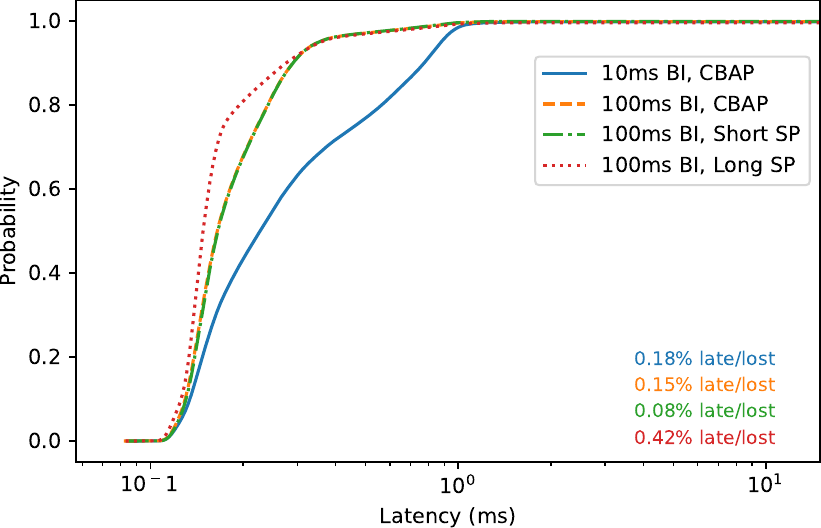}
  \caption{Repeats of well-performing configurations, with \gls{A-BFT} beam tracking}
  \label{fig:dtibf_repeat}
\end{figure}

As a final parameter, we vary where in the \gls{BI} \textbf{beam tracking} is scheduled. By default, beam tracking is scheduled during the \gls{DTI}, normally intended for data transmission. By beam tracking here, instead of during the \gls{A-BFT}, the station does not have to wait for the next \gls{BHI} to perform urgent beam tracking, for example during rapid motion. Fig.~\ref{fig:dtibf} shows that latency improves somewhat by disabling \gls{DTI} beam tracking. More importantly, packet loss is reduced by two orders of magnitude. Especially in an interactive \gls{XR} scenario, this is a massive improvement. We suspect that the firmware attempts to transmit packets during this \gls{DTI} beam tracking and does not reschedule these properly. While this is an apparent improvement, note that the experiments presented so far were for fully static environments, meaning the true impact of disabling \gls{DTI} beam tracking remains to be evaluated.

Before proceeding to the mobility experiments, we repeat a selection of the previous experiments with \gls{DTI} beam tracking disabled, with Fig.~\ref{fig:dtibf_repeat} showing the results. Differences in performance for previously evaluated parameters are now less pronounced. \SI{100}{\milli\second} \glspl{BI} perform best, with long \glspl{SP} having slightly lower packet loss. The slightly higher loss with these parameters is negligible. The lowest-loss configuration, with regular \glspl{SP}, saw \SI{99.61}{\percent} of all packets arrive within \SI{1}{\milli\second} and \SI{99.92}{\percent} within \SI{2}{\milli\second}. Less than \SI{0.0028}{\percent} of packets arrived with higher latency, and \SI{0.075}{\percent} were lost.

\subsection{Mobility Experiments}
\begin{figure}[!t]
  \centering
  \includegraphics[width=\linewidth]{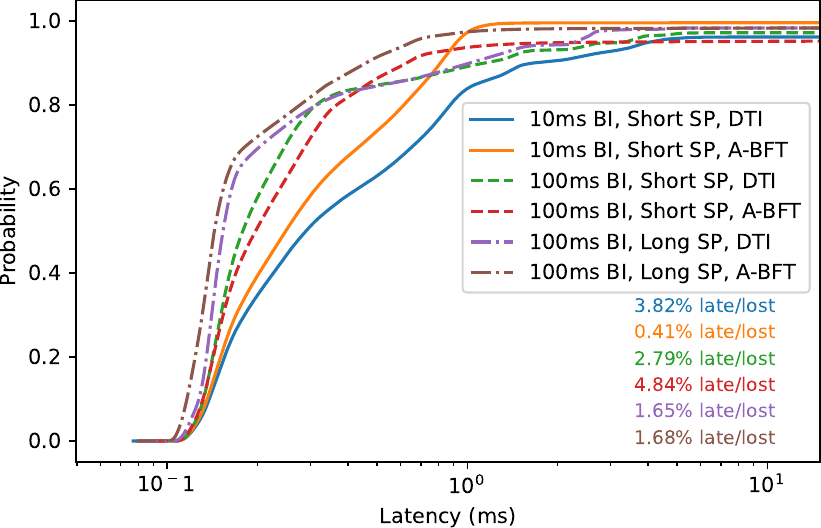}
  \caption{Latency under moderate motion}
  \label{fig:rot}
\end{figure}

\begin{figure}[!t]
  \centering
  \includegraphics[width=\linewidth]{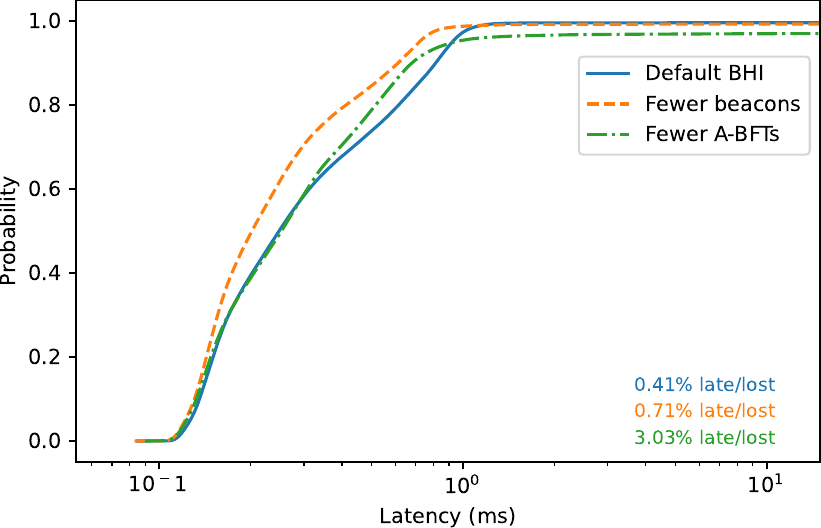}
  \caption{Latency for different \gls{BHI} configurations, under moderate motion}
  \label{fig:rot_bhi}
\end{figure}

The mobility patterns in these experiments were chosen to mimic mobile \gls{XR}, meaning the transmitter is always static. Firstly, the receiver rotates at moderate speed. The receiving \gls{EVK} was rotated continuously (alternating between clockwise and counterclockwise), completing a \SI{90}{\degree} turn every \SI{2}{\second}. The \gls{AoA} and broadside were aligned at the midpoint of the rotation, meaning the \gls{AP} stays within the \gls{XR} device's \gls{FoV}. Fig.~\ref{fig:rot} shows latencies for possibly viable configurations. Notably, a short \gls{BI} with \gls{A-BFT} beam tracking performs best, with a packet loss of only \SI{0.41}{\percent}. \gls{DTI} beam tracking is meant to speed up beam tracking, but clearly, at moderate rotational speeds, the \gls{BHI} alone can offer sufficiently responsive beam tracking. The \gls{DTI} beam tracking only hinders data transmission even in this case. At larger \gls{BI}, non-negligible loss occurs even with \gls{DTI} beam tracking (though it is difficult to tell if this occurs due to collisions or link degradation). Looking at latencies, the \SI{10}{\milli\second} configuration without \gls{DTI} beam tracking, despite its \gls{CDF} rising more slowly than others’, has the highest percentage of packets arrive rapidly, with \SI{97.32}{\percent} within \SI{1}{\milli\second} and \SI{99.55}{\percent} within \SI{2}{\milli\second}. \SI{0.037}{\percent} arrived with a higher latency, and \SI{0.41}{\percent} were lost. For the next experiments, we use a \SI{10}{\milli\second} \gls{BI} and \gls{A-BFT} beam tracking.

Then, we repeat the \gls{BHI} optimization experiments, which improved latency using a \SI{10}{\milli\second} \gls{BI} in the static case. This gives the latencies in Fig.~\ref{fig:rot_bhi}. Decreasing the number of beacons has a slight positive effect on latency but increases loss somewhat, while reducing the \gls{A-BFT} frequency has a serious impact on reliability. This occurs because the \gls{EVK} needs to wait significantly longer before beam tracking, causing queues to fill up and latencies to increase to over half a second.
\begin{figure}[!t]
  \centering
  \includegraphics[width=\linewidth]{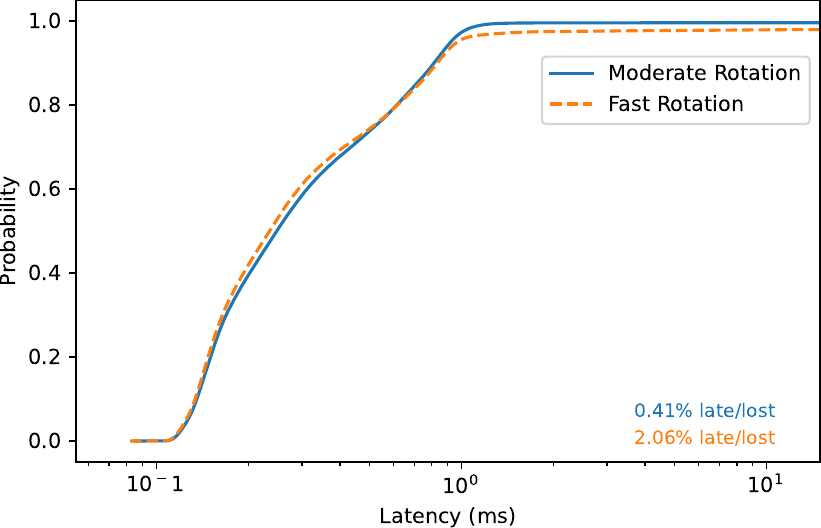}
  \caption{Latency under rapid motion}
  \label{fig:fastrot}
\end{figure}

As a final experiment with rotations within the \gls{FoV}, we consider fast rotations. We repeat the best-performing configuration, now rotating the \gls{EVK} as rapidly as safely possible every \SI{2}{\second}, leaving the \gls{EVK} in the final position until the next rotation. Fig.~\ref{fig:fastrot} shows that, while the overall latency distribution is similar, the loss increases five-fold with faster rotations, indicating the beam tracking algorithm does not always adapt in time.

\begin{figure}[!t]
  \centering
  \includegraphics[width=\linewidth]{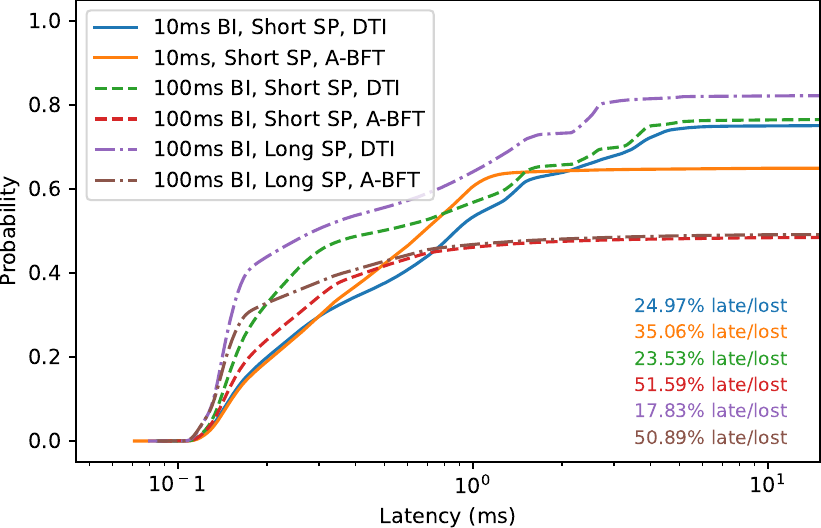}
  \caption{Latency under extreme motion}
  \label{fig:hugerot}
\end{figure}

Finally, we evaluate what happens when the \gls{AP} leaves the \gls{XR} device's \gls{FoV}. In this scenario, the device is turned \SI{180}{\degree} every \SI{2.5}{\second}, with the rotation itself taking \SI{1.5}{\second}, meaning the antenna arrays are perpendicular to each other around \SI{40}{\percent} of the time, making successful reception extremely challenging. As Fig.~\ref{fig:hugerot} shows, every considered configuration performs too poorly to be usable. While there are significant differences between the configurations, re-establishing a broken link sometimes took multiple seconds, meaning the only valid conclusion from this experiment is that more than 1 \gls{AP} is needed to reliably cover a highly mobile user.

\section{Conclusions}
In this work, we evaluated the performance of state-of-the-art IEEE 802.11ad/ay hardware. We measured latency distribution and throughput in both static and mobile scenarios, with a focus on rapid rotational motion expected in \gls{XR} scenarios. In addition, we measured the impact of modifying several PHY and \gls{MAC}-layer parameters. By tuning the structure of the \gls{BI}, we were able to increase throughput from \SI{1.85}{Gbps} to \SI{2.12}{Gbps}, a nearly \SI{15}{\percent} increase. This configuration, with a \SI{100}{\milli\second} \gls{BI}, \gls{A-BFT} beam tracking and long \glspl{SP}, exhibits a loss around \SI{0.4}{\percent} and near-consistent sub-\SI{2}{\milli\second} packet latency for arriving packets, with shorter \glspl{SP} reducing packet loss to under \SI{0.1}{\percent}, but reducing throughput by \SI{0.14}{Gbps}. Under moderate motion, we maintained a loss of around \SI{0.4}{\percent} by reducing the \gls{BI} length to \SI{10}{\milli\second}. This allowed for more frequent beam tracking, but reduced throughput back to \SI{1.85}{Gbps}. Fast motion increased this loss to \SI{2}{\percent}, while extreme motion, with the \gls{AP} leaving the \gls{XR} device's \gls{FoV}, increased loss to \SI{35}{\percent}. Overall, these experiments show that, while \gls{mmWave} is a promising enabler for mobile interactive \gls{XR}, faster or even proactive beam tracking solutions for mobile nodes are needed. Furthermore, solutions such as multi-\gls{AP} deployments, \gls{RIS}~\cite{ris} and distributed antenna systems~\cite{das} can increase coverage. Even then, the streaming protocol will need to be robust to some packet loss.
\section*{Acknowledgment}

This research was partially funded by the ICON project INTERACT and Research Foundation - Flanders (FWO) project WaveVR (Grant number G034322N). INTERACT was realized in collaboration with imec, with project support from VLAIO (Flanders Innovation and Entrepreneurship). Project partners are imec, Rhinox, Pharrowtech, Dekimo and TEO.

\bibliographystyle{IEEEtran}
\bibliography{IEEEabrv,bibliography}
\end{document}